\documentclass{aa}
\usepackage{graphics}

\newcommand{\comment}[1]{}
        \def\smallskip{\vskip 2pt}

\begin{document}

\thesaurus{06(08.05.3; 08.16.4) }

\title{Is the core mass-luminosity relation violated by the occurrence
of the third dredge-up ?}

\author{P.\ Marigo$^{1,2}$, L.\ Girardi$^{1}$, A.\ Weiss$^{1}$, 
M.A.T.\ Groenewegen$^{1}$}
\institute{
$^1$ Max-Planck-Institut f\"ur Astrophysik, Karl-Schwarzschild-Str.\
	1, D-85740 Garching bei M\"unchen, Germany \\
$^2$ Department of Astronomy, University of Padova,
        Vicolo dell'Osservatorio 5, I-35122 Padova, Italy}

\offprints{Paola Marigo \\ e-mail: marigo@pd.astro.it} 

\date{Received 6 April 1999 / Accepted 19 August 1999}

\maketitle
\markboth{P.\ Marigo et al.}{The third dredge-up and the 
$M_{\rm c}-L$ relation}

\begin{abstract}

The core mass-luminosity ($M_{\rm c}-L$) relation of thermally pulsing
asymptotic giant branch (TP-AGB) stars is a key ingredient in
synthetic calculations of their evolution. Recently, Herwig et
al.~(1998) have presented full calculations of TP-AGB models with
strong dredge-up occurring already during early thermal pulses. The
resulting luminosity evolution differs appreciably from the simple
linear $M_{\rm c}-L$ relation.

In this paper, we show that at least part of the luminosity evolution
can be understood as being the result of two well-known effects: the
gradual approach to the asymptotic behaviour that characterises the
first thermal pulses, and the chemical composition changes of the
envelope. Both effects are already implemented in the $M_{\rm c}-L$
relations used in synthetic models. Consequently, these models are
able to reproduce the behaviour of full calculations. Whether
additional effects, not yet taken into account, are present, can be
decided only through additional calculations and data. We also comment
on the validity of a linear $M_{\rm c}-L$ relation and its possible
violation, as mentioned by Herwig et al.

\comment{
Herwig et al.\ (1998) recently claimed that thermally pulsing
asymptotic giant branch (TP-AGB) stars experiencing efficient third
dredge-up do not follow the core mass-luminosity ($M_{\rm c}-L$)
relation. In this paper, we argue that the effect they found probably
does not represent a real violation of the $M_{\rm c}-L$ relation.

Various effects may be contributing to the apparently abnormal
luminosity evolution of their models. In particular, we mention (1) the
gradual approach to the asymptotic 
$M_{\rm c}-L$ relation that characterizes the
first thermal pulses, when the gravitational energy 
from the contracting core provides a significant
fraction of the stellar luminosity; and (2) the already known
composition dependence
of the $M_{\rm c}-L$ relation, which makes the quiescent luminosity
to shift to higher values than predicted for constant metallicity,
especially in models with very efficient dredge-up.  The statement by
Herwig et al.\ that synthetic models of the TP-AGB phase use linear 
core mass -- luminosity relations is incorrect, as both effects are
already included in the relations currently
used. We compare predictions
of synthetic evolution models using such $M_{\rm c}-L$ relations with
the computations of Herwig et al.\ and discuss remaining discrepancies
and their possible origin. 
}


\end{abstract}


\section{Introduction}
\label{intro} 

The $M_{\rm c}-L$ relation for TP-AGB stars, first discovered by
Paczy\'nski (1970a), has since then been employed in many studies
involving this evolutionary phase. It is a basic ingredient in
synthetic calculations of TP-AGB evolution, and an important tool for
the interpretation of observational data for AGB stars.

What the $M_{\rm c}-L$ relation means in the classical sense is
simply that the quiescent luminosity of a TP-AGB star in the
full-amplitude regime is mainly controlled by its core mass, without
any dependence on the mass of its outer envelope.

Various theoretical analyses have been performed in the past to
explain the existence of the $M_{\rm c}-L$ relation from first
principles, using either homology relations (Refsdal \& Weigert 1970;
Havazelet \& Barkat 1979; Kippenhahn 1981), or the equations of
stellar structure under specific physical conditions (see Eggleton
1967; Paczy\'nski 1970b; Tuchman et al.\ 1983; Jeffery 1988).

A transparent discussion of the validity of the $M_{\rm c}-L$ relation
was presented by Tuchman et al.\ (1983), to whom the reader should
refer. There it is shown that an $M_{\rm c}-L$ relation {\em necessarily}
holds when the star consists of:
\begin{itemize}
\item a {\it degenerate core} of mass $M_{\rm c}$ surrounded by
\item a narrow {\it radiative burning shell} (or double shell) source
providing most of the luminosity ($L_{\rm H}\sim L$), 
beyond which there must exist
\item a {\it thin } (with a mass $\Delta M \ll M_{\rm c}$) and {\it
inert} (the luminosity is constant) {\it transition region in
radiative equilibrium}, extending up to the base of the convective
envelope.
\end{itemize}
Then, because of the extreme steepness of the structural gradients
across the radiative inert zone, it follows that the thermal
evolution of the core is decoupled from that of the envelope.
The relationship between the core mass 
and the luminosity, defined on the ground of this physical picture,
is of linear nature, as confirmed by  numerical results (e.g.\ Paczy\'nski
1970a; Iben 1977; Wood \& Zarro 1981; Boothroyd \& Sackmann 1988a).
Hereinafter, such linear relation will be referred to as {\it
the classical $M_{\rm c}-L$ relation}. 

However, such a simple $M_{\rm c}-L$ relation does not hold for all
AGB stars.  Bl\"ocker \& Sch\"onberner (1991) have shown that the
$M_{\rm c}-L$ relation can indeed break down in the most massive AGB
stars ($M \ga 3.5-4.5 M_{\odot}$ depending on the metallicity)
experiencing envelope burning (or hot-bottom burning).  In this
context, substantial efforts have been made in order to accurately
include this effect in synthetic TP-AGB calculations (Marigo et al.\
1998; Marigo 1998; Wagenhuber \& Groenewegen 1998).
It must be emphasized that $M_{\rm c}-L$ relations in synthetic
calculations are always technically motivated relations intended to
fit results of full stellar evolution calculations. By no means they are
just the classical, physically motivated linear relations
mentioned above.
In the following, such relations
will be referred to as {\it technical $M_{\rm c}-L$ relations}.

Very recently, Herwig et al.\ (1998, hereinafter HSB98) have claimed
that the classical $M_{\rm c}-L$ relation may also be violated in low-mass AGB
stars, as a consequence of efficient third dredge-up. More
specifically, they present evolutionary sequences with a dredge-up
efficiency close to $\lambda=1$ or even higher\footnote{$\lambda$ is
defined as the ratio between the dredged-up mass and the core mass
increase during each inter-pulse period.}, i.e. characterized 
by an almost constant or slightly decreasing core mass $M_{\rm c}$. 
Despite of this fact, these sequences are found to evolve at increasing
luminosity. This behavior is in apparent contradiction with the
trend expected from the classical $M_{\rm c}-L$ relation, predicting 
lower luminosities at lower core masses.

In the present study we address the question whether these results
present a {\em further} deviation from the classical $M_{\rm c}-L$
relation, as claimed by HSB98, or if they can be explained, at least
partly, by 
the already known (and understood) deviations.

\section{Violations of the classical $M_{\rm c}-L$ relation}

A violation of the classical $M_{\rm c}-L$ relation implies that, for
some reason, the configuration defined in Tuchman et al.\ (1983) is
altered.  For instance, the occurrence of hot-bottom burning (or
envelope burning) in the most massive TP-AGB stars causes the inert
radiative buffer to disappear, due to the deep penetration of the
convective envelope into the H-burning shell. Another example refers
to the first inter-pulse periods, when the luminosity of a TP-AGB star
is found to be lower than predicted by the classical $M_{\rm c}-L$
relation for the same $M_{\rm c}$.  In these initial stages the
condition $L_{\rm H} \sim L$ is not actually fulfilled, as the
gravitational contraction of the core and the He-burning shell provide
non-negligible contributions to the surface luminosity.

In the context of the recent results by HSB98, the first natural
question is: does the third dredge-up in low mass TP-AGB stars lead to
a real violation of the classical $M_{\rm c}-L$ relation? 
In other words, is any
of the conditions listed above not fulfilled?

The answer is: no.  In fact, the degenerate core and the H-burning
shell still exist in the quiescent regime after the dredge-up has
occurred, as does the radiative inert buffer, since HSB98 are only
considering stars which do not experience hot-bottom burning.

The second question is: if the basic conditions for the existence of
the classical $M_{\rm c}-L$ relation are still fulfilled, what causes
the deviation from the linear $M_{\rm c}-L$ relation?

\subsection{The first pulses}
\label{sec_initpulses}

In order to answer the latter question, let us consider the previously
known deviations from the classical $M_{\rm c}-L$ relation.

The first effect to consider is the initial luminosity evolution of
TP-AGB stars. In complete calculations of AGB stars the first thermal
pulses still take place during a phase of fast core contraction, at
luminosities lower than given by the classical, linear $M_{\rm c}-L$
relation. During a few thermal pulses, the luminosity gradually
approaches this relation, up to the so-called {\it full-amplitude
regime}.  During these first pulses unique relations between $M_{\rm
c}$ and $L$ are not expected to exist.

\begin{figure}[th]
\resizebox{\hsize}{!}{\includegraphics{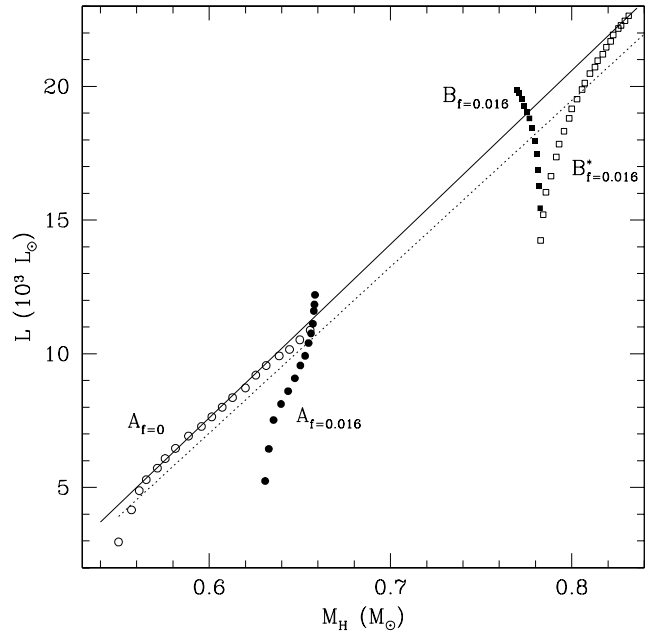}}
\caption{Evolution of the pre-flash quiescent luminosity for the HSB98
models.  The symbols refer to the inter-pulse periods, and are taken
from Fig.~2 of HSB98. The continuous line represents a linear
$M_{\rm c}-L$ relation which is chosen here to describe the full-amplitude
regime of the $A_{f=0}$ and $B^*_{f=0.016}$ sequences. The dotted line
shows the $M_{\rm c}-L$ relation of Bl\"ocker (1993).}
\label{fig_lmc}
\end{figure}

The sequences of models shown by HSB98 in their Figs.~2 and
3 refer to a relatively small number of pulse cycles, most of which
have not yet attained the full-amplitude regime. More specifically,
the two evolutionary sequences they computed with overshooting, for $3
M_{\odot}$ (labelled $A_{f=0.016}$ in HSB98) and $4 M_{\odot}$
($B_{f=0.016}$) models, present only 14 and 12 thermal pulses, respectively.
In each of these sequences, at least 6 of the pulse cycles are clearly
in the sub-luminous phase which characterizes the first pulses. This
can be seen in Fig.~\ref{fig_lmc}.

HSB98 compare their sequences with the $M_{\rm c}-L$ relation from
Bl\"ocker (1993). The latter is shown as a dotted line in
Fig.~\ref{fig_lmc}. This relation clearly predicts too faint
luminosities if compared to the most luminous points in the sequences
of models with no or little dredge-up, $A_{f=0}$ and
$B^*_{f=0.016}$. This inappropriateness of the Bl\"ocker $M_{\rm c}-L$
relation to describe the present HSB98 models probably derive from the
different input physics used in both sets of models.

We therefore prefer to define another linear $M_{\rm c}-L$ relation,
more appropriate to describe the asymptotic behaviour of the HSB98
models without dredge-up. This is shown as the solid line in
Fig.~\ref{fig_lmc}. This has been chosen to be the one which
reasonably fits the 16 last inter-pulse periods (out of 19) in the
$A_{f=0}$ sequence, and the last 3 or 4 (out of 23) in the
$B^*_{f=0.016}$ one. Both evolutionary sequences 
asymptotically approach this linear relation.

\comment{
It is now necessary to define more clearly the classical $M_{\rm c}-L$
relation which the HSB98 tracks should be compared to. The two
sequences of models with no or little dredge-up, $A_{f=0}$ and
$B^*_{f=0.016}$, seem to asymptotically approach a unique linear
$M_{\rm c}-L$ relation, the solid line in Fig.~\ref{fig_lmc}. This relation
is well defined by the 16 last inter-pulse periods (out of 19) in
$A_{f=0}$, and by the last 3 or 4 (out of 23) in $B^*_{f=0.016}$. It
is then reasonable to consider that these inter-pulse periods define
the $M_{\rm c}-L$ relation for the HSB98 set of models. HSB98 instead
compare their sequences with the $M_{\rm c}-L$ relation from Bl\"ocker
(1993), which clearly passes below the relation we draw (see
Fig.~\ref{fig_lmc}). The reason for the difference between the
two relations probably resides in the different input physics of
Bl\"ocker's models with respect to HSB98 ones. Moreover, technical details
might matter, as HSB98 trace the striking difference between sequences
$B_{f=0.016}$ and $B^*_{f=0.016}$ to details of the grid resolution in
their models.
}

In all the HSB98 evolutionary tracks, the first thermal pulses have
luminosities which are lower than predicted by this asymptotic linear
$M_{\rm c}-L$ relation we adopt (solid line). In the sequences with efficient
dredge-up, only the very last quiescent models are more luminous than 
predicted by this relation for the same core masses.  
Specifically, only the last 3 inter-pulse periods of
the $A_{f=0.016}$ track, and the last 5 of the $B_{f=0.016}$ one are
above the $M_{\rm c}-L$ relation. The remaining points are all
steadily increasing in luminosity, which is just the behavior
expected for the first thermal pulses.
 
The luminosity increase in the initial phase of the TP-AGB evolution
is partly due to the release of gravitational energy by the
contracting core, and clearly constitutes a violation of the
assumptions for the validity of the classical $M_{\rm c}-L$
relation. However, this effect is well known and already taken into
account in the technical $M_{\rm c}-L$ relations in synthetic
models (e.g.\ Groenewegen \& de Jong 1993; 
Marigo et al. 1996, 1998, 1999; Marigo 1998; Wagenhuber \& Groenewegen 1998). 
We note, however, that the behavior of $L(M_{\rm c})$ for the two
sequences with efficient dredge-up clearly deviates from those of
the sub-luminous pulses without dredge-up, such that the presence of an
additional effect resulting from the dredge-up is likely.

\subsection{The composition dependence in the $M_{\rm c}-L$ relation}

Another important point is related to the change of the surface
chemical composition produced by the third dredge-up, and thus to the
composition dependence of the $M_{\rm c}-L$ relation.

Indeed, the fact that changes in the chemical composition of the
envelope may affect -- but do not violate -- the classical $M_{\rm
c}-L$ relation had already been pointed out long ago from theoretical
arguments (e.g.\ Refsdal \& Weigert 1970; Kippenhahn 1981; Tuchman et
al.\ 1983).  As clearly derived from Tuchman et al.\ (1983; see their
equations~(1.17) and (1.29)) the $M_{\rm c}-L$ relation contains a
non-negligible dependence on the composition of the envelope,
essentially expressed by three parameters:
\begin{itemize}
\item a factor ($1+X$) from the electron scattering opacity,
\item a factor ($5X+3-Z$) from the mean molecular weight 
($\mu = 4/(5X+3-Z)$ for a fully ionized gas),
\item a factor ($X Z_{\rm CNO}$) from the hydrogen burning rate.
\end{itemize}

Since then, various $M_{\rm c}-L$ relations, both classical linear and
technical ones which include a composition dependence, have been presented
by different authors (e.g.\ Lattanzio 1986; Boothroyd \& Sackmann
1988a; Wagenhuber \& Groenewegen 1998; Tuchman \& Truran 1998).  From
these studies it turns out that at any given core mass, the quiescent
luminosity of a TP-AGB star increases with increasing metallicity $Z$,
helium content $Y$, (both leading to a higher mean molecular weight
$\mu$), and CNO abundances $Z_{\rm CNO}$.  For instance, based on
calculations of full AGB models, Boothroyd \& Sackmann (1988a)
carefully analyzed the composition dependence, deriving a proportionality
factor $\sim Z_{\rm CNO}^{1/25}\;\mu^{3}$ in their fitting formula of
the  $M_{\rm c}-L$ relation.  They found that at given core
mass, stars of solar composition ($Z=0.02,\,\mu\sim 0.62$) are $\sim
25\%$ more luminous than metal poor stars ($Z=0.001,\,\mu\sim 0.598$).

It follows that the occurrence of recurrent dredge-up episodes in
TP-AGB stars is expected to alter (not to break) even the classical
$M_{\rm c}-L$ relation, as a consequence of the increase of the mean
molecular weight in the envelope.

Of course, this effect has already been included in several synthetic
calculations (e.g.\ Groenewegen \& de Jong 1993; Marigo et al.\ 1996,
1998; Marigo 1998), where $M_{\rm c}-L$ relations with a metallicity
dependence have been adopted.  Marigo (1998) has already pointed out
that a deviation from the $M_{\rm c}-L$ relation, corresponding to
constant metallicity, can be caused by changes in the envelope
composition.

We have estimated, from the data presented in 
Herwig et al.\ (1997; hereinafter HBSE97),
Herwig (1998) and HSB98, 
the total change in the envelope composition
due to the dredge-up events. For the $3 M_\odot$ $A_{f=0.016}$ track,
the mean molecular weight $\mu$ is estimated  
to increase from 0.6314 to 0.6394 during the
TP-AGB evolution, whereas for the $4 M_\odot$ $B_{f=0.016}$ one, it
increases from 0.6304 to 0.6376. This implies that in both cases $\mu$
increases by 1.3 \% in total. Assuming  $L \propto \mu^3$
(following Boothroyd \& Sackmann 1988a), this change in
the envelope chemical composition would imply a change of 4\% 
in the luminosity predicted by the linear $M_{\rm c}-L$ relation 
for constant metallicity. This already
accounts for one-third of the luminosity increase above the $M_{\rm
c}-L$ relation drawn in Fig.~\ref{fig_lmc}.

\subsection{The presence of hot-bottom burning}
\label{hbb}
HSB98 claim that their evolutionary sequences do not present
hot-bottom burning, since their core masses are ``lower than those
associated to hot-bottom burning'' (HBB). The highest core mass in their
tracks is $M_{\rm c}=0.83 M_\odot$, whereas they consider HBB to be
present only at higher core masses.

However, the knowledge of the core mass is not enough to 
diagnose the possible occurrence of HBB. 
Several authors (Boothroyd \& Sackmann 1992;
Vassiliadis \& Wood 1993; D'Antona \& Mazzitelli 1996; Marigo 1998) 
find that the
presence of HBB, and its associated ``over-luminosity'', are sensitive
to other stellar parameters as well, as e.g.\ the envelope mass,
metallicity, mixing-length parameter, and to the details of the
convection theory. 

The latter results have been obtained by means of stellar models that
adopt canonical convection theories.  It would be interesting to quantify
whether the diffusive overshooting scheme applied by HSB98 to all
convective boundaries, may also
produce conditions favorable to HBB, i.e.\ higher temperatures at
the bottom of the convective envelope 
at lower core masses. If this were the case the
over-luminosity of the tracks may be partially ascribed to the
occurrence of (a possibly mild) HBB. In this respect, however, 
no conclusion can be drawn without additional information about 
the HBS98 tracks.

\section{The dependence on the core radius}
\label{other}

As remarked above, most of the luminosity behavior of the HSB98
tracks can be understood by means of already known effects taking
place during the TP-AGB evolution.  HBS98, however, explicitly mention
a violation of the classical $M_{\rm c}-L$ relation caused by dredge-up, and
provide an explanation for the unusual behavior of their tracks based
on the stellar core radius.

The authors find that the core radii, $R_{\rm c}$, of their TP-AGB
models follow quite different paths in the $M_{\rm c}-R_{\rm c}$
plane, depending on whether the models experience dredge-up or not.
Then, considering the apparent lack of a unique $R_{\rm c}-M_{\rm c}$
relationship and using the homology relation $L(r/R_{\rm c}) \propto
M_{\rm c}^{2}R_{\rm c}^{-1}$, HSB98 conclude that the luminosity is
not a function of $M_{\rm c}$ alone, but also of $R_{\rm c}$.  As a
consequence, the $M_{\rm c}-L$ relation should rather be seen as a
$M_{\rm c}-R_{\rm c}-L$ relation.

\begin{figure}[th]
\resizebox{\hsize}{!}{\includegraphics{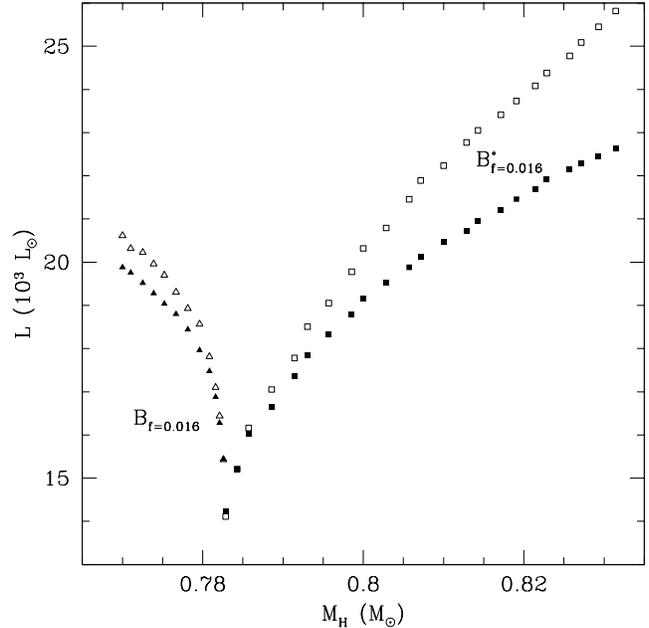}}
\caption{Evolution of the pre-flash quiescent luminosity (squares) for
the HSB98 $4 M_\odot$ models (filled dots). The open dots represent
the luminosity predicted by means of Eq.~(\protect\ref{eq_lmc2}).}
\label{fig_lmc2}
\end{figure}

Moreover, HSB98 claim that the over-luminosity above the $M_{\rm c}-L$
relation of their evolutionary sequen\-ces with $M_{\rm c}\sim0.8
M_\odot$ can be explained assuming
	\begin{equation}
L = {\rm constant}\, M_{\rm c}^{2} \, R_{\rm c}^{-1} \,\, ,
	\label{eq_lmc2}
	\end{equation}
a fact we cannot confirm. In Fig.~\ref{fig_lmc2}, we plot the
luminosity evolution of the HSB98 $4 M_\odot$ sequences, as derived
both from their $L$ and $M_{\rm c}$ values (from their Figs.~2 and 3;
filled symbols). We then use the $M_{\rm c}$ and $R_{\rm c}$ values in
order to obtain the equivalent luminosity from Eq.~(\ref{eq_lmc2})
above. This procedure however requires that we fix a value to the
constant in this equation.  The first point of the $B_{f=0.016}$
sequence is used for this purpose, so that we obtain the relation
$L=0.553\,M_{\rm c}^{2} \, R_{\rm c}^{-1}$ (where all quantities are
in solar units). The open symbols in Fig.~\ref{fig_lmc2} then show the
luminosities as obtained from this latter relation.

In this way, we have obtained the equivalent of Fig.~4 in HSB98. We
find that the luminosities as derived from Eq.~(\ref{eq_lmc2}) are far
from reproducing those given by the complete evolutionary sequences,
although the general behaviour is similar. In contrast, HSB98 obtain
quite a good match between the two sets of curves.  Examining their
Fig.~4, and comparing it with Fig.~\ref{fig_lmc2}, we conclude that
HBS98 adopt two different scales in their plot (i.e. for the true
luminosity and for the luminosity as obtained from
Eq.~(\ref{eq_lmc2})), which are not related by a single multiplicative
constant.  Thus, their explanation of the luminosity evolution in
terms of Eq.~(\ref{eq_lmc2}) is misleading, since obviously, the
``constant'' they adopt varies from model to model along the
evolutionary sequences.

The possible dependence of the luminosity on
core radius deserves the following remarks. 
This dependence would reflect the release of
gravitational energy by the contracting core. During the TP-AGB
evolution, the core contracts more rapidly during the first thermal
pulses, until an almost  constant and very low contraction rate is established
in the full-amplitude regime (Herwig 1998). 
Another concurring effect comes from the decrease
of the ratio $\beta$ between the gas and the total pressure at increasing 
luminosities. The homology relation actually predicts $L
\propto M_{\rm c}^{\sigma_1}R_{\rm c}^{\sigma_2}$, where the exponents
$\sigma_1$ and $\sigma_2$ are given in equation~3 of HSB98. Since
$\sigma_2\propto\beta$, the radius dependence in  Eq.~(\ref{eq_lmc2}) 
vanishes as we increase
$L$ and hence $\beta\rightarrow0$.

For both reasons, an $M_{\rm c}-L$ relation independent of $R_{\rm c}$ 
should hold after a
certain time. Unfortunately, the calculations by HSB98  have been stopped
at the most important point, i.e. where the 
evolution of the core radius as a function of the core 
mass for the models with efficient dredge-up joins the standard 
$M_{\rm c}-R_{\rm c}$ relation 
described by the models without dredge-up after the initial pulses 
(see their figure 3).
If, from this point on, both relations follow the same path, then the entire 
effect presented by HSB98 is indeed related to peculiar behaviour of the first
pulses, before the settling of the full-amplitude regime. In this case,
there would be no real violation of the $M_{\rm c}-L$ relation.
 
Otherwise, if there is a different dependence of the core radius 
upon the core mass, the $M_{\rm c}-L$ relation might be at least 
partly modified.
If this were the case, it would be quite important to  single out  
the physical effect produced  by 
the convective dredge-up, which occurs at a thermal pulse, 
on the core radius  during the subsequent quiescent evolution.
In other words, why should the evolution of the core radius turn out 
different between models with and without dredge-up? This point 
is not clear in the HSB98 analysis.

To this respect, an interesting point is discussed 
by Tuchman et al. (1983), in their 
analytical demonstration of the $M_{\rm c}-L$
relation from first principles.
In brief, the authors shows that core radius of an AGB star is larger than the 
radius of an ideal zero-temperature white dwarf (for which  a unique 
linear $M_{\rm c}-R_{\rm c}$ relation exists) by a multiplicative
factor, $\alpha$,  which depends both on the core mass $M_{\rm c}$ and 
on the temperature $T_{\rm c}$ at the top of the H-burning shell 
(i.e. bottom of the overlying inert radiative buffer; see their equations
1.18 and 1.19). This factor $\alpha$, being typically $\la 3$ for relevant
burning shells,  
decreases with $M_{\rm c}$ and increases with
$T_{\rm c}$. Hence, in order to get a greater shrinkage of the core
in the AGB models with efficient dredge-up, while the core mass is 
kept constant, a lower temperature
$T_{\rm c}$ should be attained during the quiescent regime.
This, in fact, would result in a smaller $\alpha$,
and hence in a smaller $R_{\rm c}$. The final result would be a certain
excess of luminosity with respect to the reference $M_{\rm c}-L$
relation. Unfortunately, no information about $T_{\rm c}$ is given in 
HSB98, but it would be worth investigating this point with the aid of full
AGB calculations. 

%
\begin{figure}[th]
\resizebox{\hsize}{!}{\includegraphics{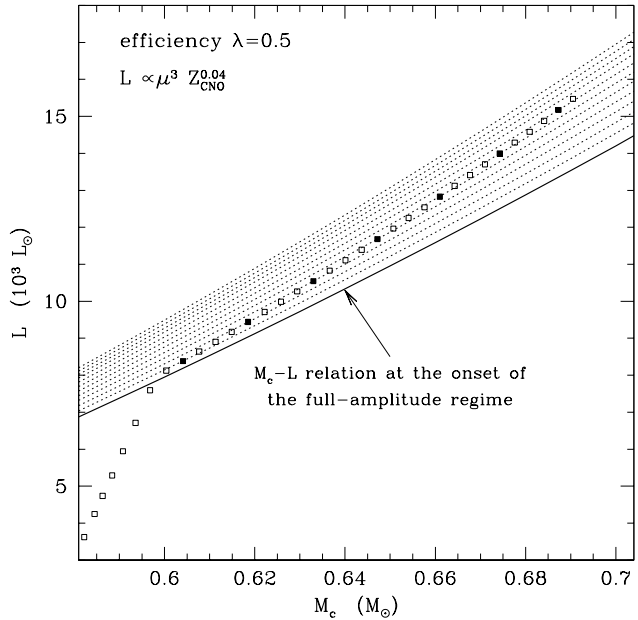}}
\caption{Evolution of the pre-flash quiescent luminosity (squares) for
the $3 M_{\odot}$ model, assuming $\lambda=0.5$, the composition of
the dredged-up material from Herwig et al.\ (1997), and the $M_{\rm
c}-L$ relation from Boothroyd \& Sackmann (1988a). The grid of dotted
lines corresponds to $M_{\rm c}-L$ relations for various values of
$\mu$ and $Z_{\rm CNO}$ (both increasing with $L$).  The filled
squares mark a few selected values of the quiescent luminosity, each 
determined, at given core mass, by that $M_{\rm c}-L$ relation of 
the grid which is consistent
with the current surface chemical composition.}
\label{lmcgrid}
\end{figure}
%

\section{Synthetic calculations with dredge-up}
\label{synt}
%
\begin{figure*}
\resizebox{12cm}{!}{\includegraphics{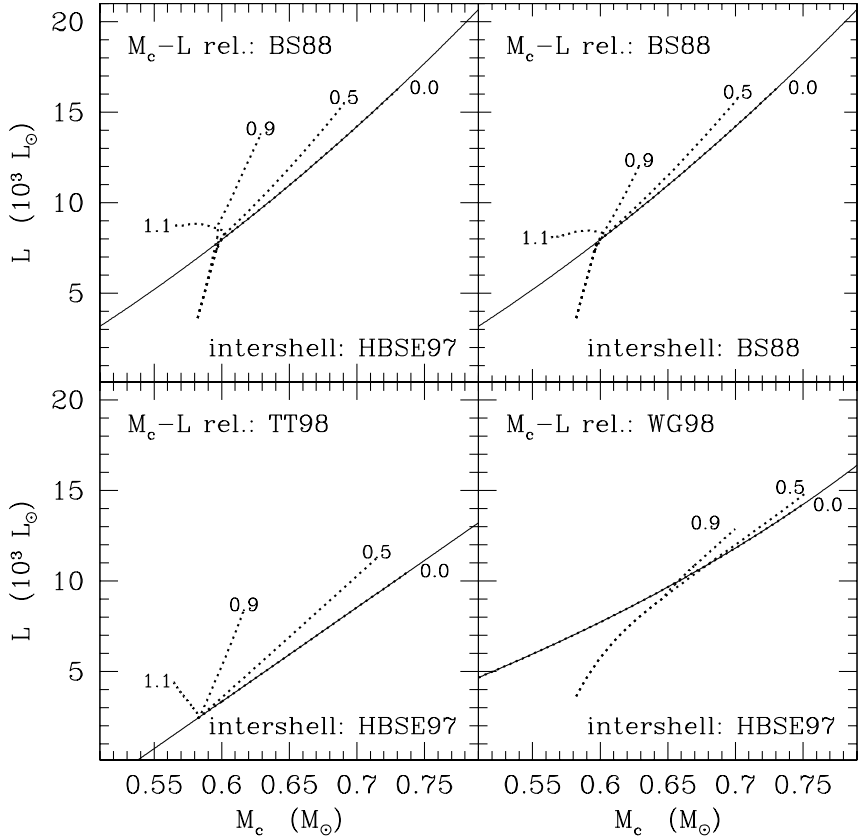}}
\hfill
\parbox[b]{55mm}{
\caption{Quiescent luminosity (i.e.\ at the pre-flash maximum) as a
function of the core mass for an evolving $3 M_{\odot}$ TP-AGB star
(dotted lines). The luminosity evolution is
derived adopting various $M_{\rm c}-L$ relations (i.e.\ Boothroyd \&
Sackmann 1988a, BS88; Tuchman \& Truran 1998, TT98; Wagenhuber \&
Groenewegen 1998, WG98), assuming two different prescriptions for the
chemical composition of the dredge-up material (BS88 or HBSE97; see
the text for more details), and different choices for the efficiency
parameter of dredge-up ($\lambda=0.0, 0.5, 0.9, 1.1$) as indicated
nearby the corresponding curve.  In each panel, the solid line refers
to the $M_{\rm c}-L$ relation consistent with the envelope composition
at the onset of the full-amplitude regime.}
\label{lmcd}
}
\end{figure*}
%
%
%
Here we present the results of synthetic
calculations carried out with different technical $M_{\rm c}-L$ relations
including a composition dependence and the first 
subluminous pulses (see Figs.~\ref{lmcgrid} and
\ref{lmcd}).  
The models are
meant to be useful experiments, giving a first hint of how the
quiescent luminosity of a TP-AGB star may behave when dredge-up events
strongly alter the envelope composition.

Calculations are carried out over a limited number of
inter-pulse periods for a 3 $M_{\odot}$ TP-AGB star
with original solar composition (i.e. $Z=0.01886$, $X=0.708$).
The chemical composition of the envelope at the first
thermal pulse is characterised by ($Z=0.01899, X = 0.68108,
\mu=0.62633, Z_{\rm CNO}=0.01357$).
The third dredge-up is assumed to occur once the full amplitude regime
is attained, i.e. after the first subluminous pulses when the  
linear $M_{\rm c}-L$ relation is approached.
We adopt various values of the dredge-up parameter
($\lambda = 0.,\, 0.5,\, 0.9,\, 1.1$) and two prescriptions for the
composition of the dredged-up material.
They are (in mass fraction):
\begin{itemize}
\item  $^{4}$He$=0.76$, $^{12}$C$=0.22$, $^{16}$O$=0.02$  \\
(according to Boothroyd \& Sackmann 1988b; BS88)
\item  $^{4}$He$=0.25$, $^{12}$C$=0.50$, $^{16}$O$=0.25$  \\
(according to HBSE97)
\end{itemize}

Here we are not concerned to give a detailed description of dredge-up
and its properties.  For instance, we assume that dredge-up events
take place at each thermal pulse in the full amplitude regime, whereas
we expect that a significant increase of the envelope metallicity
could, at a certain stage, even inhibit further occurrence of the
process by decreasing the temperature at the base of the convective
envelope during the post-flash luminosity maximum (see e.g.\ Boothroyd
\& Sackmann 1988c).

Since most available $M_{\rm c}-L$ formulae were obtained
for relatively small ranges of metallicity, usually not super-solar,
they may not give realistic results if the envelope metallicity
increases to very high values.  However, in this respect, 
the recent analysis developed by Tuchman \& Truran (1998) is relevant. 
They have
quantitatively investigated the composition influence upon the $M_{\rm
c}-L$ relation, in order to better estimate the luminosity of
classical novae, objects in which shell hydrogen burning is known to
occur in extremely metal-rich material (e.g.\ $Z=0.25$). 
At such high values of the metallicity, the corresponding $M_{\rm c}-
L$ relation is shifted to significantly higher luminosities
than predicted for solar composition.

Figure~\ref{lmcgrid} shows the locus traced by the $3 M_{\odot}$ TP-AGB
model experiencing dredge-up, adopting an efficiency $\lambda=0.5$ and
the HBSE97 prescription for the chemical composition of the
inter-shell.  A grid of classical $M_{\rm c}-L$ relations (from Boothroyd \&
Sackmann 1988a) is also plotted for increasing values of the mean molecular
weight, ranging from $\mu=0.62633$ to $\mu=0.65345$ in steps 
of about $0.0025$.  The envelope 
composition of the last calculated model ($34^{\rm th}$ pulse)
is characterised by ($\mu = 0.64408, \, Z_{\rm CNO} = 0.04489$).
For a  core mass $M_{\rm c} = 0.6905 M_{\odot}$ the quiescent  
luminosity is $\log L/L_{\odot} = 4.1896$, corresponding  
to an over-luminosity of about $14 \%$ with respect to the case  
in which the chemical composition were unchanged and equal to that
of the first thermal pulse.

Other examples are presented in Fig.~\ref{lmcd}. All models show that
the deviations from the $M_{\rm c}-L$ relation at constant metallicity
are greater for increasing $\lambda$ and/or for higher abundances of
carbon and oxygen in the inter-shell.  We can also note that models
with $\lambda = 1.1$ also evolve to luminosities above the reference
$M_{\rm c}-L$ relation, despite the effective decrease of the core
mass.

Finally, we remark that our synthetic results shown in Fig.~\ref{lmcd}
reproduce the behavior of the luminosity as found by HSB98. For
instance, the cases with $\lambda=1.1$ and $\lambda=0.5$ clearly
resemble the sequences B$_{f=0.016}$ and B$^*_{f=0.016}$ in their
Fig.~2, respectively. However, it must be specified that in our
calculations with extremely efficient dredge-up ($\lambda \sim 1$), 
a considerable over-luminosity above the $M_{\rm c}-L$ relation 
shows up after a  much larger number of dredge-up episodes
($\sim 10^2$) if compared to the results  by HSB98 ($\sim 10$).  
This difference can be partly ascribed to the fact that in our
case the onset of third dredge-up  occurs only when the full-amplitude regime 
is attained, whereas in HSB98 dredge-up takes place from the first
thermal pulses on, when other effects (in addition to the chemical
composition, see Sects.~\ref{sec_initpulses}\comment{, \ref{hbb}} 
and \ref{other}) 
are likely to play a role.  

\section{Concluding remarks}

In this paper we claim that the results from HSB98 may partially be
understood by means of the already known deviations from the classical
linear $M_{\rm c}-L$ relation.

An important effect certainly present in their evolutionary
calculations is the increase in luminosity associated with the initial
core contraction that occurs during the first thermal pulse cycles of
any TP-AGB star. This phase of rapid luminosity evolution represents a
substantial fraction of the tracks presented by HSB98. In order to
determine if dredge-up really leads to a violation of the classical
$M_{\rm c}-L$ relation, which is expected to hold for the later
evolution of AGB stars, the HSB98 evolutionary sequences should be
extended in order to include a much larger number of thermal pulses.

In fact, the most evident effects of the efficient dredge-up in HSB98
evolutionary sequences are: 
	\begin{enumerate}
	\item The small or negative changes in the core mass from
pulse to pulse, which cause the tracks to evolve almost vertically in
the $M_{\rm c}-L$ diagram, instead of along a line of increasing core
mass and luminosity.
	\item The changes in the surface chemical composition which
make their quiescent luminosity deviate from that predicted by an
$M_{\rm c}-L$ relation obtained for a constant value of metallicity.
	\end{enumerate}
None of these effects, however, implies a violation of the classical
$M_{\rm c}-L$ relation. The structural conditions for the existence of
a $M_{\rm c}-L$ relation are expected to hold only after the tracks
enter in the full-amplitude regime, as remarked above.

In this regard, we remark that the evolutionary tracks should be
compared with the $M_{\rm c}-L$ relation obtained from the current
chemical composition of the envelope, and not with those obtained from
tracks of constant metallicity. Also, the possible presence of
hot-bottom burning should be completely ruled out before we can tell
about deviations from the $M_{\rm c}-L$ relation. It would be of
particular interest, for instance, to investigate the evolution of
low-mass stars ($M\la2 M_\odot$) computed with a similar algorithm for
convection as in HSB98.

It turns out that the correct interpretation of HSB98 results requires
the analysis of additional quantities along their evolutionary tracks,
other than the core mass, luminosity, and core radius. These
quantities are: the fraction of the stellar luminosity provided by the
release of gravitational energy (necessary to identify if the
full-amplitude regime has been reached), the surface chemical
composition of the models (necessary to better quantify the deviation
from the initial $M_{\rm c}-L$ relation due to composition changes);
the luminosity provided by nuclear burning in the convective
envelope (necessary to rule out the presence of hot-bottom
burning); and the temperature $T_{\rm c}$ at the top of the H-burning
shell (useful to investigate its effect on the factor $\alpha$,
defined in Sect.~\ref{other}, and hence on the core radius $R_{\rm c}$). 
Unfortunately, this information is not provided by HSB98.

We stress once more that synthetic TP-AGB models have already been
adopting {\em technical non-linear} $M_{\rm c}-L$ relations, i.e.\
including significant deviations from linearity due to the
sub-luminous first thermal pulses and changes in the surface chemical
composition produced by dredge-up (e.g.\ Groenewegen \& de Jong 1993;
Marigo et al.\ 1996; Marigo 1998). Moreover, the real breakdown of the
$M_{\rm c}-L$ relation caused by hot-bottom burning in the most
massive AGB stars have been accurately taken into account (Marigo et al.\
1998; Marigo 1998; Wagenhuber \& Groenewegen 1998) in these
models. Finally, we recall that the $M_{\rm c}-L$ relation applies
only to the quiescent inter-pulse periods, but not to the luminosity
variations driven by thermal pulses.  Even the effect of the
post-flash low-luminosity dip is usually included in synthetic TP-AGB
calculations.

Therefore, synthetic AGB evolution calculations already include all
known effects affecting the $L(M_{\rm c})$-relation and do not rely on
the assumption that the classical, linear $M_{\rm c}-L$ relation is
valid. A corresponding comment in HSB98 turns out to be inappropriate. 
As such, any {\em new} effect, as possibly indicated by the
HSB98 calculations can easily be incorporated after sufficient
data from full calculations are available.

\comment{
An important aspect of synthetic TP-AGB models is that they clearly
indicate the need of efficient third dredge-up in low-mass stars (down
to $\sim1.2 M_\odot$), in order to reproduce the carbon star
luminosity functions in the Magellanic Clouds (Groenewegen \& de Jong
1993; Marigo et al.\ 1996, 1998). The best results have been obtained
with values of $\lambda\la0.7$. In these conditions, the TP-AGB
tracks have a monotonically increasing core mass, which evolution is
probably well described by present $M_{\rm c}-L$ relations.
}


\begin{acknowledgements}
We thank F.\ Herwig for providing additional information about his
results, and P.\ Wood for useful discussions. Our referee, Y. Tuchman,
is acknowledged for important comments about this paper.   
The work by L.\ Girardi is funded by
the Alexander von Humboldt-Stiftung.
\end{acknowledgements}

\end{document}